\documentclass[
  journal=pasa,
  manuscript=research-paper, %% or "review"
  year=2021,
  volume=x,
]{cup-journal}

\usepackage{microtype,siunitx,booktabs}
\usepackage{amsmath, amssymb}
\usepackage{hyperref}
\hypersetup{colorlinks,citecolor=blue,linkcolor=blue,urlcolor=blue}

\defcitealias{Ord2019}{Paper~I}

\title{MWA Tied-Array Processing IV: A Multi-Pixel Beamformer for Pulsar Surveys and Ionospheric Corrected Localisation }
\author{N. A. Swainston}
\affiliation{International Centre for Radio Astronomy Research, Curtin University, Bentley, WA 6102, Australia}
\author{N. D. R. Bhat}
\affiliation{International Centre for Radio Astronomy Research, Curtin University, Bentley, WA 6102, Australia}
\author{I. S. Morrison}
\affiliation{International Centre for Radio Astronomy Research, Curtin University, Bentley, WA 6102, Australia}
\author{S. J. McSweeney}
\affiliation{International Centre for Radio Astronomy Research, Curtin University, Bentley, WA 6102, Australia}
\author{S. M. Ord}
\affiliation{CSIRO Astronomy and Space Science, PO Box 76, Epping, NSW 1710, Australia}
\author{S. E. Tremblay}
\affiliation[0000-0001-7662-2576]{International Centre for Radio Astronomy Research, Curtin University, Bentley, WA 6102, Australia}
\author{M. Sokolowski}
\affiliation{International Centre for Radio Astronomy Research, Curtin University, Bentley, WA 6102, Australia}

\doi{10.1017/pas.\the\year.xxx}

\received {24 Nov 2021}
\revised  {17 Jan 2022}
\accepted {18 Mar 2022}
\published{}


\begin{document}

% git test
\begin{abstract}
The Murchison Widefield Array (MWA) is a low-frequency aperture array capable of high-time and frequency resolution astronomy applications such as pulsar studies.
The large field-of-view of the MWA (hundreds of square degrees) can also be exploited to attain fast survey speeds for all-sky pulsar search applications, but to maximise sensitivity requires forming thousands of tied-array beams from each voltage-capture observation.
The necessity of using calibration solutions that are separated from the target observation both temporally and spatially makes pulsar observations vulnerable to uncorrected, frequency-dependent positional offsets due to the ionosphere. These offsets may be large enough to move the source away from the centre of the tied-array beam, incurring sensitivity drops of $\sim$30-50\% in Phase II extended array configuration.
We analyse these offsets in pulsar observations and develop a method for mitigating them, improving both the source position accuracy and the sensitivity.
This analysis prompted the development of a multi-pixel beamforming functionality that can generate dozens of tied-array beams simultaneously, which runs a factor of ten times faster compared to the original single-pixel version. This enhancement makes it feasible to observe multiple pulsars within the vast field of view of the MWA and supports the ongoing large-scale pulsar survey efforts with the MWA.
We explore the extent to which ionospheric offset correction will be necessary for the MWA Phase III and the low-frequency Square Kilometre Array (SKA-Low).
\end{abstract}

\section{Introduction}
The Murchison Widefield Array (MWA) was initially built as a low-frequency connected element interferometer of 128 aperture array `tiles' consisting of 16 dual-polarisation dipole antennas \citep{Tingay2013}. This Phase I MWA was designed to be an imaging telescope to support a wide range of science, from continuum imaging of galactic and extragalactic radio sources to detecting the epoch of re-ionisation \citep{Bowman15}. One of the significant strengths of the MWA is its huge field-of-view (FoV); each tile (i.e. 4 x 4 array of dual-polarisation antennas) provides an FoV in the range from $\sim$300 to $\sim$1000 square degrees depending on the observing frequency within its 70-300 MHz operating range. This large FoV makes the MWA a highly efficient survey instrument.

As the coarse 0.5 second time resolution achievable with the MWA's hybrid correlator (Ord et al. 2015) was not adequate to support pulsar observations, a new functionality called the Voltage Capture System (VCS, \cite{tremblay15}) was developed. It allows recording channelised voltage data after the second stage of the polyphase filter bank in the MWA's signal path, providing a native time resolution of 100 $\mu$s and a frequency resolution of 10 kHz. However, recording these voltages results in very high data rates, $\sim$ 28 TB per hour, limiting the maximum possible observing time to 90 minutes due to available disk storage. These raw antenna voltages can then be calibrated and combined into a single, channelised, dual-polarisation voltage, tied-array, \textit{pencil beam} through software beamforming, as detailed in \cite{Ord2019} (hereafter, \citetalias{Ord2019}). This tied-array beamforming is essential to support high-sensitivity pulsar and fast transient science with the MWA.

The combination of the VCS and the software beamformer have been leveraged to conduct low-frequency pulsar science \citep{Meyers2017, Bhat2018, Mcsweeney2017}. This is, in part, due to the constant development of the beamformer, such as the polarimetric verification performed by \cite{Xue2019}. In \cite{Mcsweeney2020} a polyphase synthesis filter was implemented to recover some of the time resolution at the expense of frequency resolution ($\sim0.8 \mu$s and 1.28 MHz). This enabled the low-frequency range of the MWA to be exploited to obtain accurate dispersion measure measurements of millisecond pulsars \citep{Kaur2019}.

The MWA was upgraded with a further 128 tiles, extending its maximum baseline to $\sim$6 km \citep{Wayth2018}. However, the signal path remains the same, and as a result, only data from 128 of the 256 tiles can be correlated or recorded in the VCS mode at a given time. This ``Phase II'' MWA \citep{Wayth2018} can be configured either as a compact array with baselines within $\sim$300 metres or as an extended array with baselines up to $\sim$6 km.  The compact array configuration, which provides a sensitivity equivalent to that of Phase I MWA for tied-array beam processing, has a much broader beam size (a FWHM of $\sim 23^{\prime}$ at 155 MHz), almost a factor of $\sim$6 times larger than the Phase I array and hence a beam area that is $\sim$40 times larger. The larger beam size of the compact array means a smaller number of beams are needed to cover a given area within the FoV, which is more appealing for large scale pulsar surveys.

The Phase II extended array's $\sim$6 km maximum baseline is ideal for localising pulsar candidates (see Bhat et al., in prep) but, at the low frequencies of the MWA, the smaller tied-array beam size can become comparable to positional offsets due to the refractive ionosphere. This refraction is due to spatial variations of the total electron content (TEC), the lowest order of which is a slope across the FoV whose net effect on the apparent source positions can be described by a single ``bulk offset''. The positional offsets remaining after the bulk offset has been removed (e.g. during calibration) are due to higher-order variations in the TEC and are termed ``residual offsets'' in this work. Any offset not corrected for during calibration can potentially degrade the sensitivity of a detection if the offset is an appreciable fraction of the size of the tied-array beam.

\begin{figure*}[t]
\begin{center}
\includegraphics[width=\linewidth]{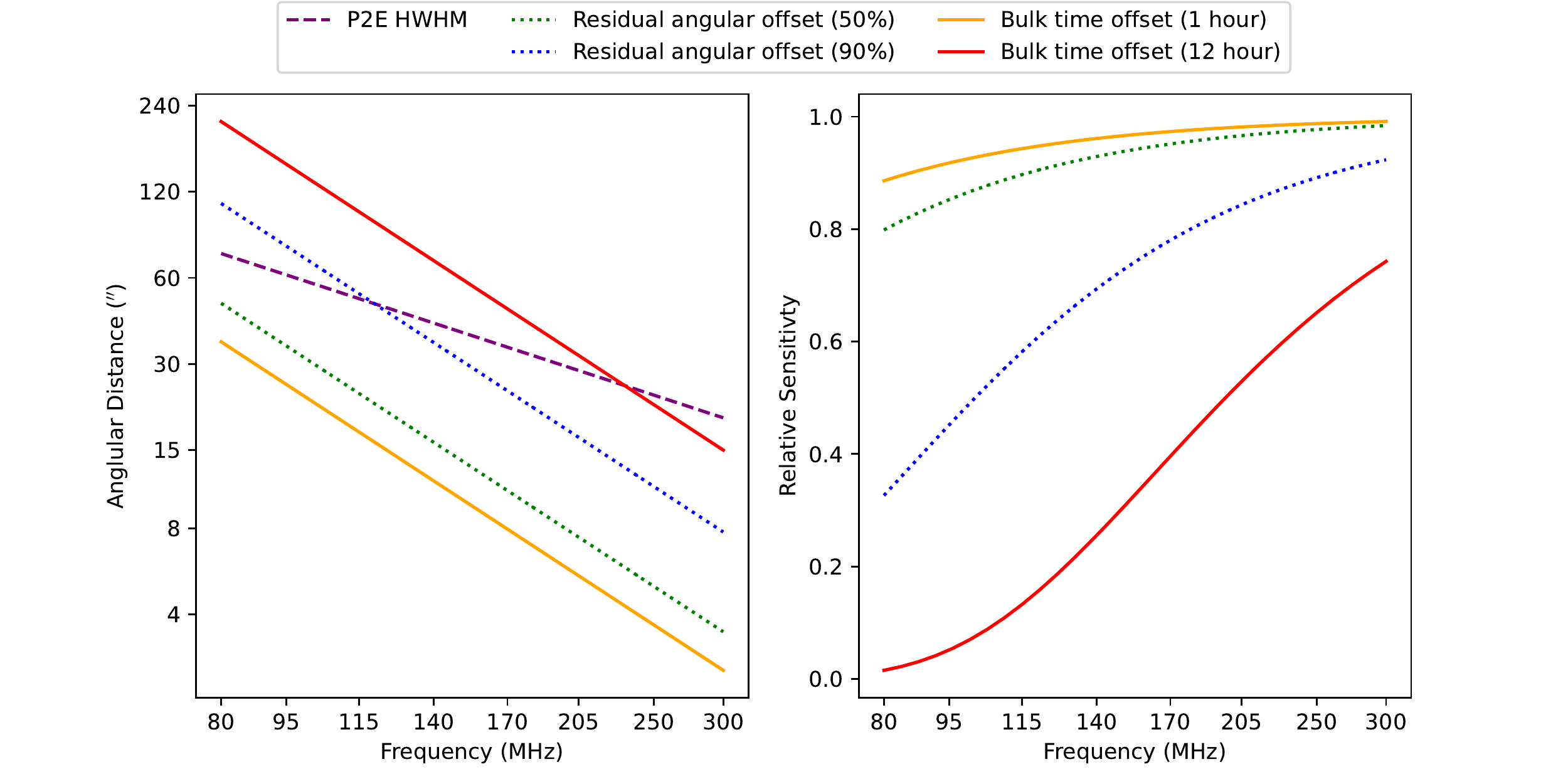}
\caption{\textit{Left}: how the MWA Phase II extended array (P2E) tied-array beam's half-width half-maximum (HWHM) scales with frequency assuming a Gaussian beam shape compared to different ionospheric offset estimates. \textit{Right}: how these offsets would affect the relative sensitivity of an observation. The ionospheric offset estimates include: Residual angular offset (50\%), the median residual offset for the 50th percentile of the observations \citep{Jordan2017}; Residual angular offset (90\%), as above but for the 90th percentile; Bulk time offset (1 hour), the maximum change in the bulk offset over 1 hour seen in the work of \cite{Arora2015}; Bulk time offset (12 hours), as above but over a 12 hour period.}
\label{fig:theory_offset}
\end{center}
\end{figure*}

Calibration of MWA VCS data typically involves an observation of a bright source and the \textit{Real Time System} (RTS, \citeauthor{Mitchell2008} \citeyear{Mitchell2008}) to create a direction independent calibration solution which is subsequently applied to the target observation. If the bulk offset is correctly accounted for, the median residual offsets are less than 0.13 $^{\prime}$ for 50\% of observations and less than 0.29  $^{\prime}$ for 90\% of observations at 200 MHz \citep{Jordan2017}. Figure \ref{fig:theory_offset} illustrates how the size of the residual offsets scales with frequency compared to the Half Width Half Maximum (HWHM). This suggests that the residuals only significantly affect sensitivity at low frequencies ($\lesssim$140 MHz), due to the offsets of the order of the beam size, for $\sim$10\% of our observations.

For the majority of MWA observations, the best calibration source is in a different part of the sky and observed at a different time than the target observation. That is, the calibration and target observations generally sample different ionospheres. Thus, the bulk offset determined during calibration can differ from the bulk offset present in the target observation. Using GPS satellites to probe the ionosphere above the MWA, \cite{Arora2015} observed the bulk offset change by up to $\sim0.17^{\prime}$ (at 150 MHz) in an hour or up to $\sim1^{\prime}$ (at 150 MHz) over 12 hours. It is standard practice to have at least one calibration observation within an hour of the target observation. However, sometimes these calibration solutions fail to converge, and we are obliged to use calibration observations up to 48 hours away. In these cases, the bulk offset has likely changed, as illustrated by the 12-hour angular distance offset in Figure \ref{fig:theory_offset}. For example, if applying the correction for an incorrect bulk offset moves a source to the half-power point of our tied-array beam, it would require observing for four times as long to recover the lost sensitivity. To prevent this, we must understand this bulk offset error and develop a method for mitigating it.

We must correct for these offsets to efficiently perform high-sensitivity pulsar and transient science in the extended array configuration and with the upcoming MWA Phase III. Our strategy for measuring and correcting positional offsets is conceptually simple: we form multiple beams around the target position and measure the strength of the detection as a function of the pointing position.
Even for the compact configuration, surveys of known pulsars may require forming up to hundreds of beams, while blind surveys require thousands of beams to tile the entire FoV, creating a processing bottleneck.
These use cases have prompted the development of new functionality that allows processing VCS data for generating multiple beams efficiently to overcome the problems associated with the VCS's high data rates. This new multi-pixel beamformer functionality will allow us to exploit the enhanced sensitivity achieved via a tied-array beam as well as the large FoV.

Similar multi-pixel beamformers have been developed to support pulsar science using other low-frequency radio telescopes such as the Giant Metrewave Radio Telescope (GMRT) \citep{Roy2012} and the Low-Frequency Array (LOFAR) \citep{Broekema2018, Sanidas2019}, both of which perform real-time processing of antenna voltages to generate $\sim$10-100 beams. The MWA beamformer, on the other hand, is conceptually different in design (Ord et al. 2019) and employs post-processing offline to generate tied-array beams. The multi-pixel beamformer functionality presented in this paper can output hundreds of tied-array beams simultaneously to make large scale pulsar surveys computationally feasible, allowing us to study and correct for ionospheric offsets.

The remainder of this paper is organised as follows. We first discuss the implementation of the MWA tied-array beamforming in \S \ref{beamforming theory}, and its upgrade to the multi-pixel beamformer in \S \ref{sec: multi-pixel beamformer}. Then in \S \ref{benchmarking}, we benchmark the improvement in processing efficiency compared to the previous beamformer using multiple supercomputers. In \S \ref{sec: applications} we demonstrate the multi-pixel beamformer's capability to correct for ionospheric offsets and perform a pulsar census. Finally, In \S \ref{sec: discussion} we discuss the implications for MWA Phase III and low-frequency Square Kilometre Array (SKA-Low).

\section{Implementation and Benchmarking}

\subsection{Tied-Array Beamforming with the MWA }\label{beamforming theory}
The design philosophy and algorithmic implementation behind tied-array beamforming with the MWA are explained in detail in \citetalias{Ord2019}
. In the following sections, we give an overview of these algorithms.

%\subsection{Delay Compensation} \label{sec: delays}
%To phase up the telescope, we must compensate for the geometric and cable delays $\Delta t$ to a single reference point. The phase correction for a channel, $n$ with centre frequency $f_n$ of antenna $j$ required to compensate for beam steering and cable delays is
%\begin{equation}
%    \phi_{j,n} = 2 \pi \Delta t_j f_n
%\end{equation}
%As the earth rotates, it alters the array's projected baselines which causes the interferometer fringe pattern on the sky to change. We can measure how the fringes change with time, known as the fringe rate, to recalculate the delays frequently enough to maximise sensitivity. Given the maximum frequency of 300 MHz and the maximum baseline of 5 km, the maximum fringe rate for the MWA Phase II is $\sim$1 rad s$^{-1}$; thus, the delays will be recalculated every second.

\subsubsection{Calibration} \label{sec: calibration}
Each antenna in the array has a complex gain, imparting a phase turn on the incoming electric field. This phase turn serves to decorrelate the sum of the antenna signals and must be compensated for so that they are on the same relative, or absolute, amplitude and phase scale. The gain calibration process is an attempt to determine the instrumental response.

Due to the antennas' lack of calibrated noise diodes, the antennas cannot be calibrated individually and must be calibrated as an interferometer. The most common method is via a short calibration scan performed on a nearby calibrator field. The raw voltages are correlated to form visibilities from which the calibration solution (i.e. complex gain of each individual antenna and polarisation) can be obtained.

These antenna-based complex gains can be described using the Jones matrix formalism \citep{Hamaker1996,Sault1996,Hamaker19962,Hamaker2000}. The Jones matrix $J_j$ for each antenna, $j$, is the complex gain that affects the incident electric field vector, $e$ that results in the measured antenna voltage
\begin{equation}
    v_j = J_j e.
\end{equation}
The RTS \citep{Mitchell2008} is a software calibrator that can be run offline to produce an estimation of the complex gains. This is done by iteratively removing residual visibilities and attempting to correct for ionospheric offsets, starting with the brightest sources.
While the RTS can correct for direction-dependent ionospheric offsets, this information is not applicable when the calibrator observation is in a different part of the sky than the target observation, as is usually the case for VCS observations\footnote{Although in-field calibration can be attempted using correlated VCS data, it can often fail to converge on a calibration solution e.g. due to the lack of bright sources in the field.}. For this reason, it is standard practice to obtain a calibration solution from a dedicated observation of a bright source.
We use this direction independent calibration solution at low radio frequencies, which incorporates the bulk ionospheric shift into the gain solutions but does not correct for residual ionospheric offsets.

%TODO expand this and talk about direction independence

\subsubsection{Beam Formation} \label{sec: beamform}
The calculation of the detected beam ($e^\prime$) is described by expanding Equation (34) from \citetalias{Ord2019}. Neglecting the noise term, for each available frequency channel,
\begin{equation} \label{eqn: beamform}
    e^\prime = \sum^{N_A}_j v_j J^{-1}_j exp \{ -i \phi_{j} \}
\end{equation}
where $v_j$ is the complex voltage from each tile, $J^{-1}_j$ is the inverse of the complex gain of the direction independent calibration estimated by the RTS (and including the primary beam correction), $exp \{ -i \phi_{j,n} \}$ is the geometric delay compensation and $N_A$ is the number of tiles (for Phase II, $N_A = 128$). This detected beam is calculated for both polarisations then transformed to the four Stokes parameters \citepalias[see Equations (47)-(59) in][]{Ord2019}.

\subsection{Multi-Pixel Beamforming Functionality} \label{sec: multi-pixel beamformer}

\begin{figure}[t]
\begin{center}
\includegraphics[width=\columnwidth]{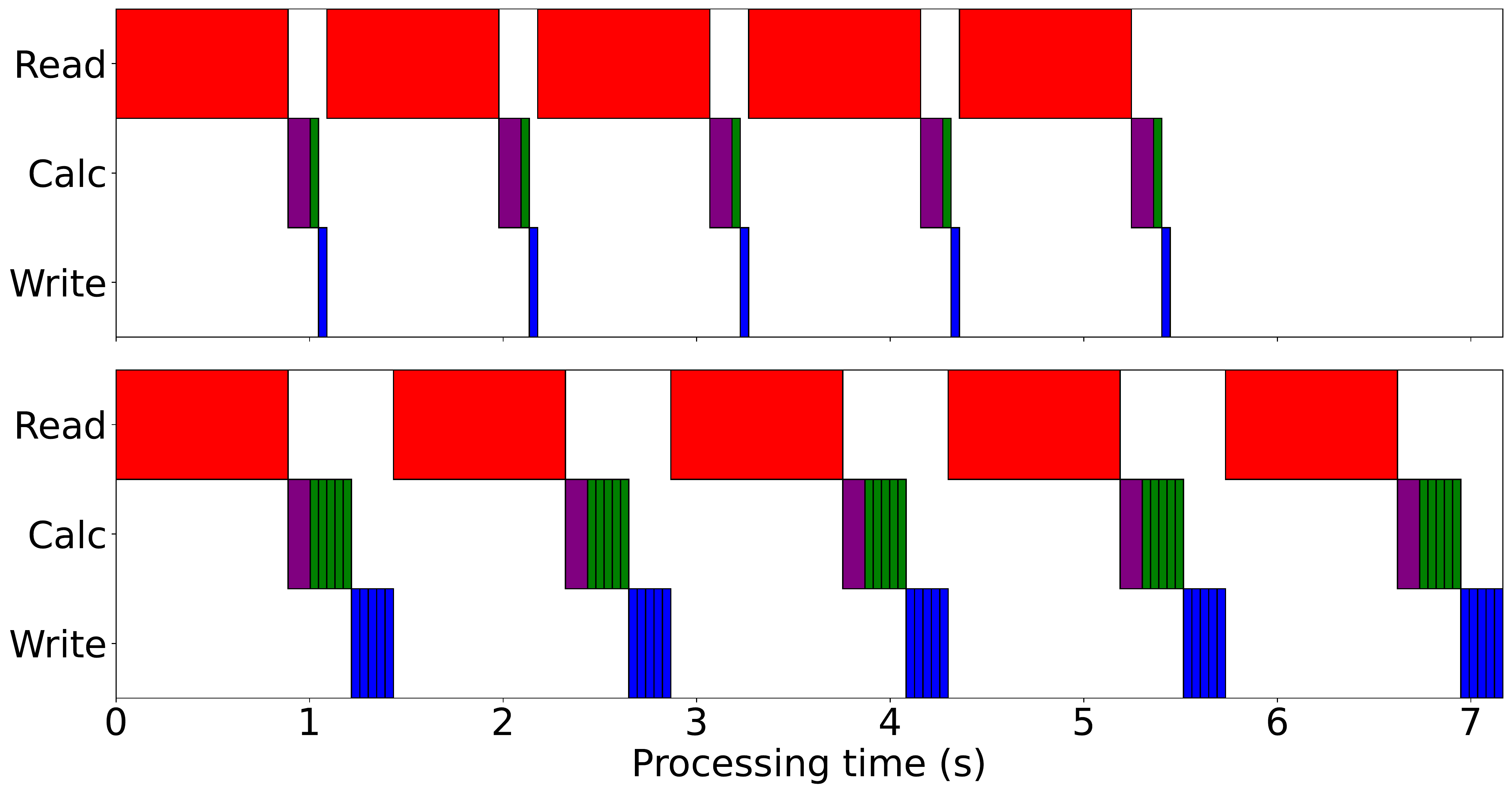}
\caption{A comparison of the processing time and workflow of the single-pixel beamformer (top panel) processing a single beam and the multi-pixel beamformer (bottom panel) processing five beams. Each block represents the processing time required to read in a second of data (red), apply the calibration solution (purple), perform the delay compensation and tile summation (green) and write out the results (blue).}
\label{fig:mpwf}
\end{center}
\end{figure}

To form multiple beams, a naive approach would be simply to repeat the calculation of Equation \eqref{eqn: beamform} for each desired pointing in the FoV.
%This is computationally expensive due to the large number of voltages, $v_j$, involved in the calculation.
%However, the calculation of the Jones matrix, $J_j$, and the geometric delays, $\exp\{-i\phi_{j,n}\}$, are only performed once per second, as these quantities are not expected to change on that time scale significantly.
This is computationally expensive due to the large size of the voltages, $v_j$, requiring significant read time, see Figure \ref{fig:mpwf}. Once these voltages have been calibrated, they can be used to beamform anywhere within the FoV.
This suggests a strategy for forming multiple beams efficiently since the only quantity that changes for different tied-array beam pointings is the geometric delay.
The geometric delay changes at a rate of 1.0 rad s$^{-1}$ which equates to a 1\% S/N drop if calculated once per second. One can therefore compute the quantities
\begin{equation} \label{eqn: cal}
    e_j = J_j^{-1} v_j
\end{equation}
just once per second of data, leaving only the summation over tiles,
\begin{equation} \label{eqn: sum beamform}
    e^\prime = \sum^{N_A}_j e_j exp \{ -i \phi_{j} \},
\end{equation}
to be performed for each desired pointing.

The complete set of computational steps required for an efficient multi-pixel beamformer would therefore be as follows (the colours listed for each step refer to the diagram shown in Figure \ref{fig:mpwf}):
\begin{enumerate}
    \item Reading in the raw MWA complex voltages (red)
    \item Applying the calibration solution and primary beam correction as per Equation \eqref{eqn: cal} (purple)
    \item Correcting for the geometric and cable delays as per Equation \eqref{eqn: sum beamform}, flattening the bandpass, and converting to Stokes parameters (green)
    \item Writing the beamformed data to disk (blue)
\end{enumerate}
Further optimisation by means of overlapping read/write tasks will be explored in the future.

We use an MWA antenna beam model to compensate for the dipole response as part of the gain compensation. The original beamformer used an analytical beam model \citepalias[described briefly in][]{Ord2019}, but since \href{https://github.com/CIRA-Pulsars-and-Transients-Group/vcstools/releases/tag/v2.3}{version 2.3} the multi-pixel beamformer has used the Full Embedded Element (FEE) primary beam model for both calibration and beamforming \citep{Sokolowski2017}. The FEE beam model simulates every dipole in the MWA tile (4 $\times$ 4 bow-tie dipoles) separately, taking into account all mutual coupling, ground screen and soil effects, and has been tested empirically by various authors \citep[e.g.][]{Line2018,Chokshi2021}. Dead dipoles, which also affect the beam response, are also taken into account during beamforming.

\subsubsection{Implementation}

The MWA beamformer has been developed as part of the \href{https://github.com/CIRA-Pulsars-and-Transients-Group/vcstools}{vcstools} repository \citep{Swainston2020}. The beamformer processes one second of data at a time, recalculating the geometric delays on the same cadence. Each second of MWA VCS data has 10,000 time samples, 3,072 frequency channels, 128 tiles and two polarisations, which equates to 8 million independent calculations of $v_j J^{-1}_j exp \{ -i \phi_{j,n} \}$ per second per tied-array beam. This computation is spread over 24 Graphics Processing Units (GPUs), one for each of the 24 coarse frequency channels. Per second of observation, this lowers the size of the input baseband voltages to 313 MB and the number of calculations to 325,000 per GPU.

%The GPUs are utilised more effectively as they spend comparatively less time idle while data is read on to the memory. %Due to the large size of the recorded antenna voltages, a large percentage of the processing time is spent reading in and calibrating these voltages instead of beamforming. Since a tied-array beam can point anywhere within the tile's FoV, the antenna voltages can instead be read in and calibrated once and used to calculate several tied-array beams, as shown in the lower panel of Figure \ref{fig:mpwf}. This increases the efficiency of the multi-pixel beamformer, per beam, as the number of beams calculated at once increases.

The large size of the raw voltages is not only a problem for processing efficiency: if running a large number of beamforming jobs, the significant demands on file I/O can affect the health of supercomputing clusters' metadata servers. These metadata servers can only handle a certain transfer rate from the file system (e.g. Lustre, which is often used on supercomputing clusters) to memory which becomes a limiting factor for large scale beamforming jobs and another reason to process as many simultaneous beams as possible with the multi-pixel beamformer.
%The geometric delays are calculated and moved on to the GPU along with the Jones Matrix and antenna voltages.

The original single-pixel implementation of the beamformer involved a GPU kernel for each beam involving the calculation shown in Equation \eqref{eqn: beamform}. This kernel was split in two so that the calculation in Equation \eqref{eqn: cal} can be performed only once per run, after which a kernel for calculating Equation \eqref{eqn: sum beamform} can be performed for each beam. Storing these intermittent products (calibrated voltages) on the GPU device may exhaust the available device memory. To prevent the GPU from running out of device memory, the beamformer automatically calculates the largest fraction of a second that can be accommodated on the GPU and processes the data in batches. Asynchronous streams were implemented to ensure that each chunk of data was moved onto the GPU and processed as soon as possible. With the above implementation in place, the only limit on the number of beams that can be processed at once is the maximum job wall time imposed by supercomputing clusters. If requested, the Stokes parameters are calculated, and the bandpass of each frequency channel flattened to account for the frequency-dependent sensitivity of the receivers. Finally, the Stokes parameters are moved off GPU memory to an output buffer.

The Stokes parameters are written to disk in either the PSRFITS \citep{Hotan2004} or VDIF \citep{Whitney2009} formats. For PSRFITS (the format used in our search pipeline), the beamformer outputs Stokes parameters to one output file per pencil beam. As the number of beams increases, opening and writing to a large number of files place an extra burden on the file system as it puts strain on the metadata servers. To prevent this, it is recommended to use Solid State Drives (SSDs), if available, to ensure writing the files does not become a bottleneck.

Blind pulsar searches using traditional search algorithms only use Stokes I. Writing only Stokes I reduces the output by a factor of 4 and can improve the efficiency of the beamformer even more than implied by the benchmarks presented in the following sections. Any pulsar candidates found using Stokes I can be re-beamformed at a later date using the full Stokes parameters for polarisation analysis.

\subsection{Benchmarking} \label{benchmarking}

The processing requirements of an all-sky pulsar search are notoriously massive and often take several years. For this reason, it is crucial to understand the beamforming bottlenecks so we can choose the supercomputer clusters that can most efficiently process the data.

\subsubsection{Relative speed improvements}

The following equation can model the efficiency improvement of the multi-pixel beamformer over the single-pixel beamformer:
\begin{equation}\label{eqn: factor}
    R = \frac{N_B (t_R + t_{C} + t_{B} + t_W)}{t_R + t_{C} + N_B( t_{B} +t_W) },
\end{equation}
where $N_B$ is the number of tied-array beams calculated at once, $t_R$ is the time it takes to read in data, $t_C$ is the time to transfer data onto the GPU and apply the complex gains, $t_B$ is the time to form the beam and calculate the Stokes parameters, and $t_W$ is the time to write the data to disk. This theoretical prediction of the improvement is compared to the benchmarked improvement in Table \ref{table:Benchmarks_parts} and illustrated in Figure \ref{fig:Benchmarks}.

%The efficiency of the multi-pixel beamformer increases as the number of simultaneous beams being processed increases, as illustrated in Figure \ref{fig:Benchmarks}.
%The improvement in processing per beam approaches an asymptote at around 15 beams as $t_R + t_C << N_B ( t_B + t_W)$ (see Equation \ref{eqn: factor}).

\subsubsection{Supercomputer platforms}

To ensure the processing load of a pulsar survey can be spread between multiple supercomputers and that all collaborators can process VCS data, we made our software portable enough to be easily installed on multiple supercomputing clusters through containerisation\footnote{\href {https://hub.docker.com/repository/docker/cirapulsarsandtransients/vcstools}{vcstools DockerHub}}.

The beamformer was initially installed and developed on the Pawsey Supercomputing Centre's Galaxy\footnote{\href{https://support.pawsey.org.au/documentation/display/US/HPC+Systems\#HPCSystems-Galaxy(CrayXC30)}{Pawsey Supercomputing Centre's Galaxy supercomputer}} supercomputer, which is used to support Australian Square Kilometre Array Pathfinder (ASKAP) and the MWA's radio astronomy processing. However, all MWA processing at Pawsey has since been migrated to the new Garrawarla cluster\footnote{\href{https://support.pawsey.org.au/documentation/display/US/Garrawarla+Documentation}{Pawsey Supercomputing Centre's Garrawarla supercomputer}}, and the beamformer software is now installed and running on that system. To spread the processing load, we additionally installed our beamforming software on Swinburne University's OzSTAR supercomputer\footnote{\href{https://supercomputing.swin.edu.au/ozstar/}{Swinburne University's OzSTAR supercomputer}} and China SKA Regional Centre's (CSRC) prototype supercomputer. Benchmarks are presented for all three systems.

%Using Equation \ref{eqn: factor}, we can estimate a theoretical improvement of the multi-pixel beamformer given the benchmarks of each part of the code, as shown in Table \ref{table:Benchmarks_compare}. This theoretical improvement is compared in Table \ref{table:Benchmarks_compare} to the measured benchmarks of the multi-pixel beamformer processing 15 beams simultaneously on each supercomputer.

\begin{table*}[t]
\begin{center}
\begin{tabular}{llrrrrrrrr}
\hline
Super      & GPU               & TFLOP & N$_{GPU}$ & $t_R$ & $t_C$ & $t_B$ & $t_W$ & $F_T$& $F_B$ \\
computer   &                   &       &           &  (ms) &  (ms) &  (ms) &  (ms) &      &       \\
\hline \hline
Garrawarla & NVIDIA V100 PCIE  & 7.0   &  78       &   677 &    80 &    33 &    20 & 8.9 & 7.7  \\
OzSTAR     & NVIDIA P100 PCIe  & 4.7   & 214       &   266 &   117 &    42 &    42 & 8.0 & 10.4 \\
CSRC       & NVIDIA V100 SXM   & 7.8   &  16       &  1329 &    36 &    54 &    32 & 9.3 & 8.4  \\
\hline
\end{tabular}
\caption{The benchmarks of each part of the MWA multi-pixel beamformer on three supercomputers where GPU is the brand/model of graphics card, TFLOPS (TeraFlops) is the peak performance for double precision of the graphics card, N$_{GPU}$ is the total number of GPUs available on the supercomputer. The following are estimates of the time required to process a second of data at each step where $t_R$ is the time spent reading in data, $t_C$ is the time spent transferring data onto the GPU and applying the complex gains, $t_B$ is the time spent forming the beam and calculating the Stokes parameters and $t_W$ is the time spent writing the data to disk. There are also factors of improved processing efficiency for 20 beams where $F_T$ is the theoretical improvement using Equation \ref{eqn: factor} and $F_B$ is the measured improvement from benchmarking.}
\label{table:Benchmarks_parts}
\end{center}
\end{table*}

\subsubsection{Benchmarking method}

\begin{figure}[t]
\begin{center}
\includegraphics[width=\columnwidth]{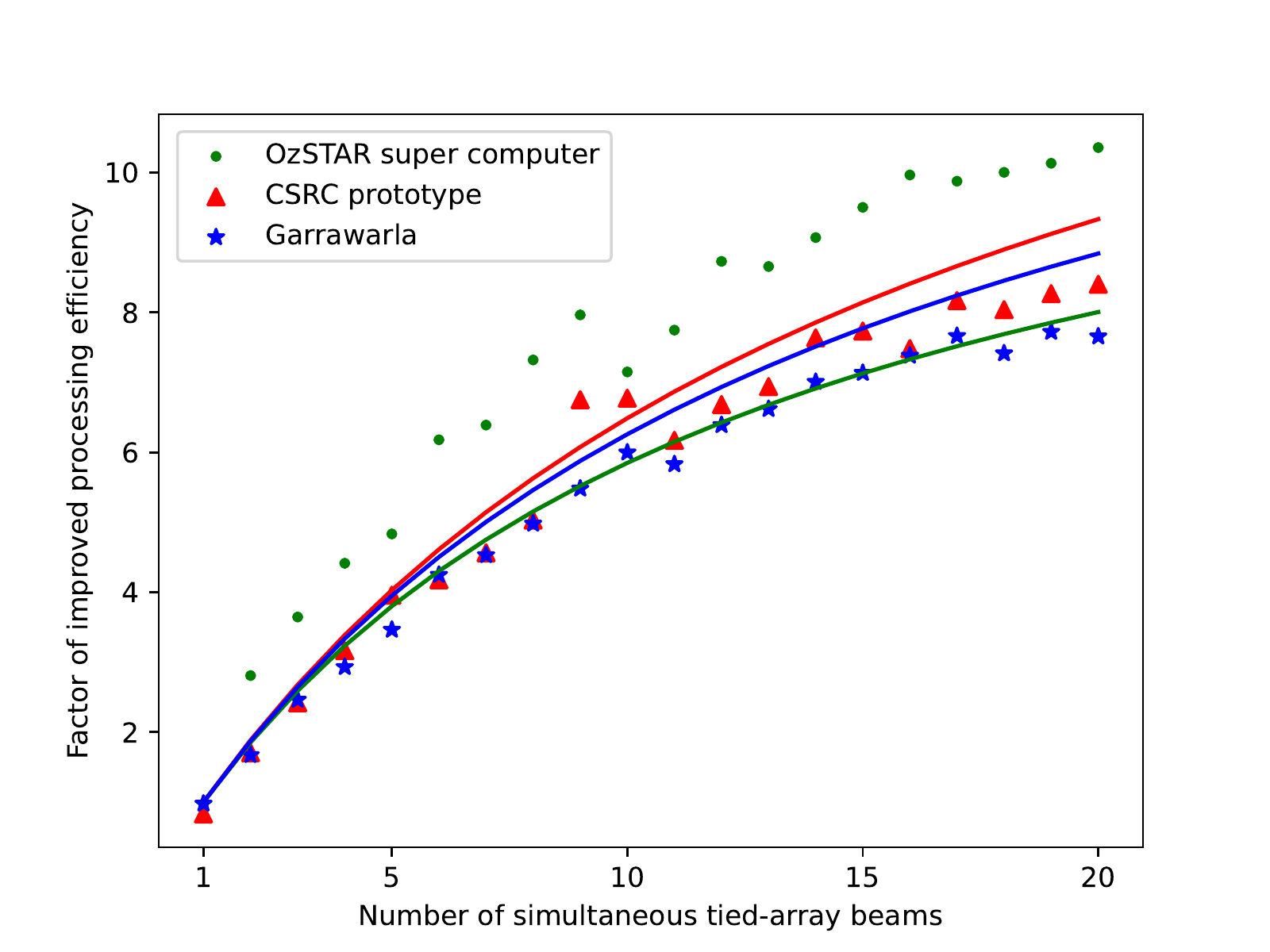}
\caption{A comparison of the processing efficiency improvement of the multi-pixel beamformer for a given number of beams on the OzSTAR (green), China SKA Regional Centre's prototype (red) and Garrawarla (blue) supercomputers. The processing efficiency per tied-array beam is an increasing function of the number of simultaneously calculated beams for the multi-pixel beamformer.}
\label{fig:Benchmarks}
\end{center}
\end{figure}

The read/input and write/output speeds can fluctuate due to how much strain the supercomputer's metadata server is under at any given time. To account for this fluctuation, the multi-pixel beamformer was benchmarked by running 24 10-minute instances using 1 to 20 simultaneous beams and compared to the single-pixel beamformer. This fluctuation still exists, leaving a $\sim$10\% variability on all read and write benchmarks. The improvement is illustrated in Figure \ref{fig:Benchmarks} and agrees with our improvement prediction.

At 20 simultaneous beams, the improvement of the multi-pixel beamformer is a factor of 7.7, 10.4 and 8.4 compared to the single-pixel beamformer for Garrawarla, OzSTAR and the CSRC prototype, respectively (see Table \ref{table:Benchmarks_compare}). Once $t_R < NB \times (t_B + T_W)$, the beamformer is no longer limited by the time spent reading in the data, and the new limiting factor becomes the time spent on the GPU and writing to disk. Thus, technological improvements such as faster GPUs and the use of SSDs can significantly improve the beamformer's processing rate.

In Table \ref{table:Benchmarks_parts} we compare the processing required to tile a 10-minute observation with (MWA Phase II compact array) tied-array beams for both the single and multi-pixel beamformers. Performing a pulsar search with the MWA requires a large number of dispersion trials to maintain sensitivity due to the increased dispersion effect at our low-frequency range. Therefore to do even a simple periodic pulsar search on the 6000 beams would require approximately 20 thousand CPU hours, which is similar to the GPU hours required by the single-pixel beamformer, see Table \ref{table:Benchmarks_compare}. The multi-pixel beamformer only takes a tenth of the processing time, meaning beamforming is no longer a bottleneck, and a blind pulsar search with the MWA is feasible.

%OzSTAR has the greatest improvement due to $t_R + t_C >> t_B + t_W$ unlike Galaxy and CSRC that have $t_R + t_C > t_B + t_W$. Since for OzSTAR $t_R + t_C >> t_B + t_W$ the single-pixel beamformer spent a larger percentage of its time reading in data so, as the improvement starts to asymtote ($t_R + t_C << N_B (0.9 t_B + t_W)$), the increase is more significant.

\begin{table}[t]
\begin{center}
\begin{tabular}{llccc}
\hline
Super      & Software  &  1B     &  20B   & 6000 Beams \\
Computer   &  version  &  (s)    &  (s)   &  (kSU)     \\
\hline \hline
Garrawarla & SPB       & \;\;479 &    479 &     19.1 \\
Garrawarla & MPB       & \;\;490 & \;\;63 & \;\; 2.5 \\
OzSTAR     & SPB       & \;\;973 &    973 &     38.8 \\
OzSTAR     & MPB       & \;\;999 & \;\;94 & \;\; 3.7 \\
CSRC       & SPB       & \;\;884 &    884 &     35.4 \\
CSRC       & MPB       &    1064 &    105 & \;\; 4.2 \\
\hline
\end{tabular}
\caption{A comparison of the original single-pixel beamformer (SPB) and the multi-pixel beamformer (MPB) processing times in seconds per tied-array beam per coarse frequency channel for a 10-minute MWA observation where 1 B and 20 B represent calculating 1 beam or 20 beams simultaneously.
``6000 Beams'' indicates the processing time in kSU (thousand service units) to process the $\sim$6000 tied-array beams required to tile the entire FoV of a 10-minute MWA Phase II compact array observation.}
\label{table:Benchmarks_compare}
\end{center}
\end{table}

\section{Applications} \label{sec: applications}

\begin{figure*}[t]
\begin{center}
%[width=\columnwidth]
\includegraphics[width=\linewidth]{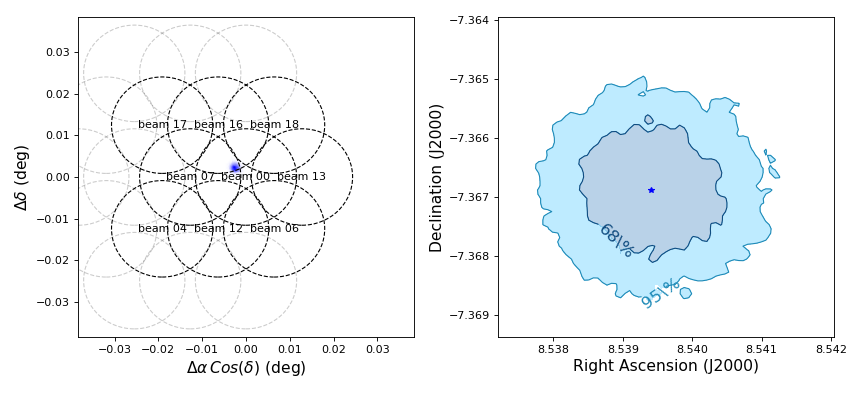}
\caption{The localisation of PSR J0036-1033 \citep{Swainston2021a} in observation 1292933216. The observation is centred at 155\,MHz and the tied-array beam has a FWHM of $\sim$1.26$^{\prime}$. The localisation method is as described in \cite{Bannister2017} and estimates the position to within 14$^{\prime\prime}$. \textit{Left}: the dashed lines represent the FWHM of each of the tied-array beams. The faint grey dashed lines are beams that are more than a beam width away and therefore not included in the localisation calculation. \textit{Right}: the first (dark blue) and second (light blue) confidence intervals of the localisation.}
\label{fig:position_estimate}
\end{center}
\end{figure*}

The improved efficiency of the multi-pixel beamformer makes large scale processing such as pulsar surveys and candidate localisation \citep{Swainston2021a} computationally feasible. Unlike other telescopes, the MWA VCS can beamform in post-processing and create a grid of pointings to estimate the position of the source without the need for re-observation. This allows the MWA to quickly localise candidates for follow up and since the beams are simultaneous, they are in the same RFI environment, so the signal-to-noise ratios of the detections can be used as a reliable proxy for comparative sensitivity. We use the localisation method as described in \cite{Bannister2017} and shown in Figure \ref{fig:position_estimate}. This method provides a beam localisation uncertainty $\sigma_L = 14^{\prime\prime}$, but this does not take into account any errors in calibration and residual ionospheric offsets. The method for minimising the calibration errors is explained in the following sections.

\subsection{Testing the validity of calibration solutions and ionospheric corrections}
When the MWA is in the extended array configuration, the FWHM ($\sim$1.26$^{\prime}$ at 155 MHz) of the tied-array beams is small enough to affect detection sensitivity when the ionosphere moves the apparent position of sources. The RTS will attempt to correct for the bulk ionospheric offset estimated by the calibration solution. If the calibration source is observed at a different time and in a different part of the sky, the ionosphere may change, leading to a different bulk offset (see Figure \ref{fig:theory_offset}).

\begin{figure*}[t]
\begin{center}
%[width=\columnwidth]
\includegraphics[width=\linewidth]{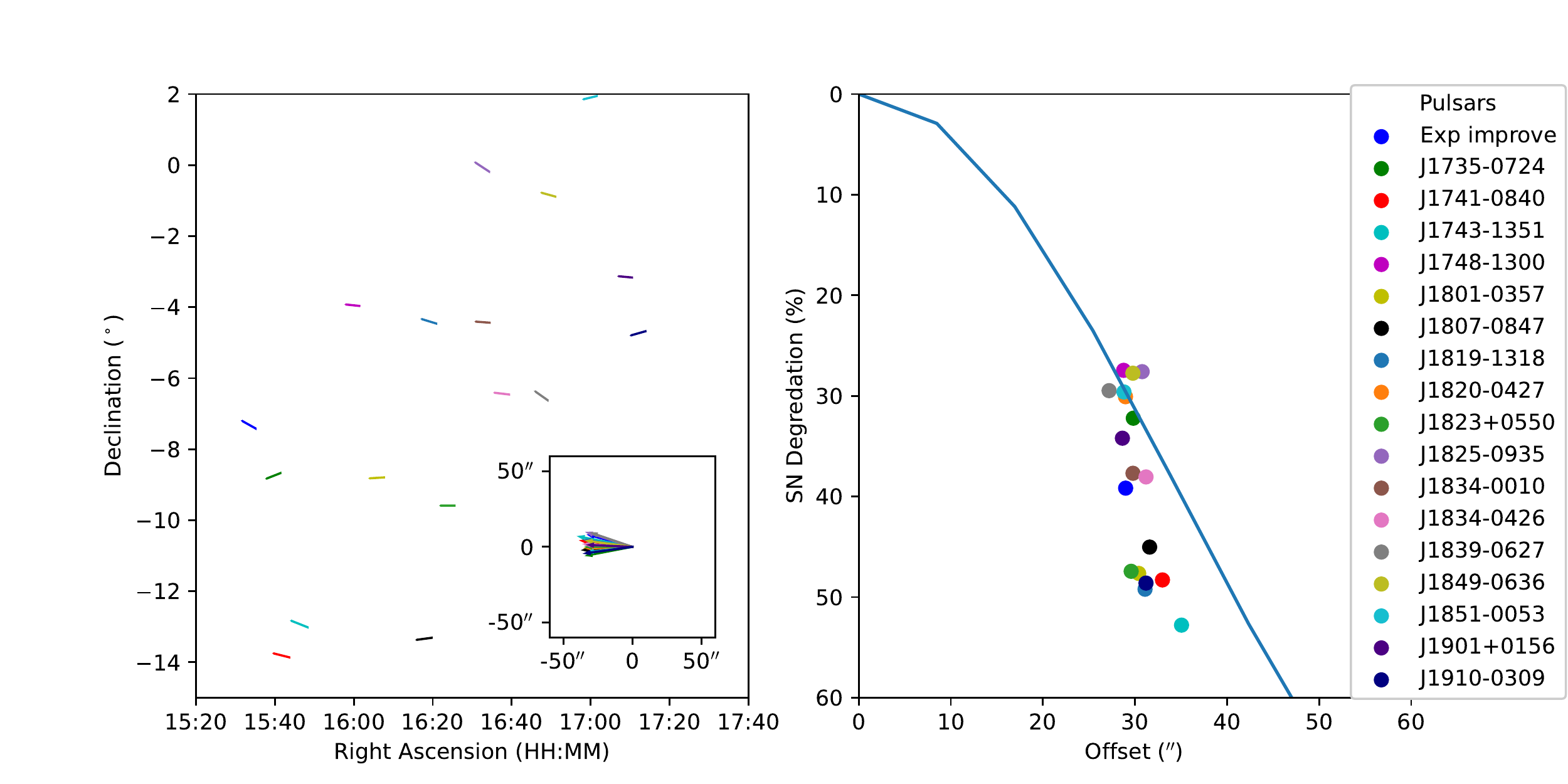}
\caption{The results of position estimation using a grid of pointings around the 18 pulsars in observation 1276619416. $\textit{left inset}$: The difference between the position estimated using the method shown in Figure \ref{fig:position_estimate} and the known position from the ATNF pulsar catalogue. $\textit{left}$: The offsets of each pulsar increased by a factor of 100, so they are visible for each pulsar's position to show that after subtracting the bulk offset, there does not appear to be any obvious correlation between the direction of the residual offsets and sky position. $\textit{right}$: The degradation in the signal-to-noise ratio (SN) of the pulsar due to its incorrect position and its total offset. The blue line represents the expected degradation using the naturally weighted point spread function generated by taking the Fourier transform of the projected baselines.}
\label{fig:bulkoffset}
\end{center}
\end{figure*}

We used the 18 pulsars detected in observation 1276619416 (taken on MJD 59019; as part of the G0071 project to study pulsar emission physics) to investigate the offsets that the ionosphere causes. These offsets were calculated by making a grid of pointings around the catalogue positions, as shown in Figure \ref{fig:position_estimate}, and estimating its position based on the measured signal-to-noise ratios. The left plot of Figure \ref{fig:bulkoffset} shows that offsets cause the sources to appear $\sim$30 $^{\prime\prime}$ away in a single direction that is independent of where they are in the FoV. We, therefore, believe this offset is caused by incorrectly accounting for the bulk offset.

%The degradation of the signal-to-noise ratio of the pulsar at the catalogue position compared to the apparent position is shown in the right plot of Figure \ref{fig:bulkoffset}.
%This is compared to the expected degradation in signal-to-noise ratio using the FWHM given from the point spread function and assuming a Gaussian beam.
We compared this to the theoretical degradation in the signal-to-noise ratio that would arise due to the offsets placing the targets significantly far from the centre of the tied-array beam.
The applied beamforming operation \citep{Ord2019} is equivalent to summing each baseline with equal weighting. In imaging parlance, this is the same as applying a "natural" weighting.
We estimated the beam response with the naturally weighted point spread function, generated by taking the Fourier transform of the projected baselines, which in this case was produced when imaging the data with the WSCLEAN software \citep{Offringa2014}.
Taking a 1D (horizontal) cut through the point spread function produces a theoretical sensitivity curve as a function of offset, which is shown in the right plot of Figure \ref{fig:bulkoffset}.
There is good agreement between the measured and theoretical signal-to-noise ratio degradation, but the slightly steeper slope of the measured points suggests that our beam has more sensitivity close to the centre of the beam, leading to a sharper fall-off.

If such an offset is not corrected for, an observation's duration would have to increase by a factor of $\sim$3 to recover the $\sim$40\% loss in sensitivity. Because there are often hundreds of known pulsars in an observation's FoV, it is inefficient to create a grid of tied-array beams around every pulsar to correct for any offsets. Instead, we have developed an efficient method for measuring and correcting for an incorrect bulk offset which is explained in the next section. After correcting the bulk ionospheric offset, the residual ionospheric offsets will remain, which cannot be corrected without direction-dependent calibration on the same field. As the residual offsets are typically $<10^{\prime\prime}$ \citep{Jordan2017}, they only cause $<5\%$ reduction in detection sensitivity, which is factored into our position estimate uncertainties.

\subsubsection{Correcting for incorrect bulk ionospheric offsets} \label{sec:offsetfix}
Any residual bulk ionospheric offsets must be measured and removed to ensure maximum sensitivity and accuracy of pulsar position estimates. To measure the offset, we choose at least three bright pulsars (with a signal-to-noise ratio above 20) within the FoV, and form a grid of pointings around them to estimate each pulsar's apparent position (see Figure \ref{fig:position_estimate}). To ensure that the average of the measured offsets most accurately reflects the bulk offset (instead of a localised ionospheric disturbance), we preferentially choose pulsars that are spread as widely as possible across the FoV. If three suitable pulsars cannot be found within the FoV, pulsars with a lower S/N can also be used. The offset between the apparent and true pulsar positions is averaged for the three pulsars, and this average becomes our estimate for the bulk offset. This bulk offset is subtracted from the position of subsequent pulsar detections in that observation to derive position estimates.

After removing the bulk offset, the uncertainty of the derived position will be dominated by the residual offsets $\sigma_R$. Although the bulk offset derived in this way will only be an approximation of the true bulk offset, the uncertainty of the bulk offset $\sigma_B$ will be significantly smaller than the average residual offsets as long as the selected pulsars sample spatially independent ionospheric shifts. However, as noted above, this assumption can fail if the chosen pulsars are too close to each other on the sky, or if there are large scale spatial structures in the ionosphere across the primary beam. For the purposes of estimating the positional errors, we assume that the measured ionospheric shifts are not biased in this way. Thus, the position uncertainty, $\sigma_P$, is the quadrature sum of the standard deviation of the magnitudes of the residual offsets, $\sigma_R$, and the localisation uncertainty $\sigma_L$:

\begin{equation}
    \sigma_P = \sqrt{\sigma_R^2 + \sigma_L^2}.
\end{equation}

\subsection{Detection of known pulsars within a field-of-view}

The MWA has already performed a pulsar census \citep{Xue2017} using the incoherent beam, which preserves the entire single tile FoV ($\sim$610 deg$^2$ at 150 MHz) but is a factor of $\sim$10 less sensitive than the tied-array beam. To perform an MWA tied-array beam pulsar census, we must create a tied-array beam on each known pulsar with a dispersion measure below 250 pc cm$^{-3}$ within the field-of-view. Because there are often hundreds of known pulsars in an observation's FoV, we use the bulk offset correction method described in \S \ref{sec:offsetfix} to efficiently ensure the maximum sensitivity.

%TODO add how many we detected
To demonstrate the effectiveness of the multi-pixel beamformer and the bulk offset correction method, we beamformed on the 256 pulsars in the FoV of observation 1276619416 and detected 18 pulsars (see Figure \ref{fig:bulkoffset}). Observation 1276619416 is in Phase II extended array configuration, has a centre frequency of 184.96 MHz and a tied-array beam FWHM of 1.05$^\prime$. Thanks to the bulk offset correction method, the signal-to-noise ratio of these detections improved by $\sim$30-50\%.

The original single-pixel beamformer was then used to reprocess all the pulsar detections in observation 1276619416 to compare the signal-to-noise ratio with the multi-pixel beamformer detections. The difference in signal-to-noise ratio is, on average, less than 1\% and is likely due to floating-point rounding errors. These results validate that a multi-pixel beamformer is required to process observations efficiently and can do so with equal sensitivity to the original single-pixel beamformer.

\section{Discussion} \label{sec: discussion}

\subsection{Survey feasibility}
The required GPU time to create tied-array beams for the entire FoV for a 10-minute observation using the Garrawarla, OzSTAR and CSRC supercomputers are shown in Table \ref{table:Benchmarks_parts}. Using OzSTAR benchmarks from Table \ref{table:Benchmarks_compare} and assuming an average of 16 GPUs are available for our use, we can approximate how much wall time it would take to process each 10-minute observation. This comes to about 14 weeks for the single-pixel beamformer but only 22 days with the multi-pixel beamformer. This equates to 19 years of processing to create the $\sim$700,000 tied-array beams required to cover the Southern Sky with the single-pixel beamformer but only $\sim$1.8 years with the multi-pixel beamformer. This enhancement dramatically improves the feasibility of performing a Southern Sky survey with the MWA.

\subsection{Implications for MWA Phase III and SKA-Low}

The Phase II MWA extended array has a maximum baseline of 6 km and a tied-array beam HWHM of $\sim40^{\prime\prime}$ (at $155\,$MHz), which is small enough to be potentially affected by the ionospheric effects described in this paper.
The imminent upgrade of the MWA to Phase III will allow all 256 tiles to be correlated and recorded simultaneously. This will include the same 6 km baselines of the Phase II extended array, so all future observations will have to consider minimising or mitigating these ionospheric offsets. The ionospheric residual offsets will begin to cause sensitivity loss for observations below $\sim$ 140 MHz when there is high ionospheric turbulence, as indicated in Figure \ref{fig:theory_offset}.

The stochastic nature of the ionosphere means we cannot predict, even to first order, how variations will behave over time or in different parts of the sky. The effect this turbulence has on the bulk offset over time was observed in \cite{Arora2015} and shown to change by up to $\sim$0.17$^{\prime}$ (at 150 MHz) in an hour with no observable patterns. How the ionosphere behaves in different parts of the sky has not been studied, but we can assume that there could be variations of $\sim$0.5$^{\prime}$.
Since we frequently use calibration observations over 50 degrees away from the target observation, we may be correcting for a different bulk offset. However, we cannot predict which calibration observations will cause an incorrect bulk offset correction as there is no clear correlation with time from observation or distance from the target observation position. For example, one calibration solution obtained from an observation separated by 31 hours yielded no significant bulk offset when applied to the target observation. On the other hand, the example shown in Figure \ref{fig:bulkoffset} uses a calibration observation with in an hour and demonstrates a $\sim30^{\prime\prime}$ offset which led to a $\sim40\%$ reduction in sensitivity. Therefore, using the bulk offset correction method for all observations with calibration solutions more than an hour away is recommended.
%\citep{Wayth2018}
%talk about the default operation mode

This has implications for real-time beamforming systems, which are desirable given the increased data rate of the Phase III high-time resolution (HTR) observing mode. The current VCS delivers (4+4)-bit complex samples for 128 dual-polarisation tiles at a $\sim$ 28 TB/hour data rate. In contrast, Phase III \citep{Wayth2018} will deliver (8+8)-bit complex samples for 256 tiles, which will quadruple the data rate to $\sim$ 112 TB/hour. This increased data rate will make real-time beamforming more desirable as these tile voltages will not have to be stored or transferred for post-processing. However, these ionospheric offsets are more problematic for real-time beamforming since they cannot be corrected in post-processing.

%The archiving of these voltages allows for flexible post-processing but is not necessary for all science cases. Simple monitoring of a single source, for pulsar timing, for example, would benefit from a real-time beamformer that would significantly reduce the data rate.

%ORIGINAL TEXT (Nick)
%Telescopes like the GMRT often observe in several sub-arrays to reduce their maximum baseline to ensure the detection of sources with high position uncertainties \citep{Gupta2017}. Similarly, using sub-arrays is a viable strategy for SKA-Low to ensure that the FWHM is wide enough to mitigate the ionospheric effects discussed in this work.

% Revised text (RB)
Besides the MWA, both LOFAR and uGMRT are two other prominent low-frequency facilities that operate at $\sim$100-200 MHz band, with baselines extending out to $\sim$10 km or longer. While LOFAR offers a substantial collecting area within a $\sim$1 km baseline, phased-array observations with the uGMRT may need to employ antennas located well outside the central square for higher sensitivity. Even though the uGMRT Band 2 (120-240 MHz) is not the most preferred observing band due to RFI considerations, it is still an order of magnitude more sensitive compared to the Phase 3 MWA, provided the signals from far-arm antennas (up to $\sim$25 km baselines) can be coherently combined. While the sub-array capabilities of the uGMRT can be exploited for mitigating potential ionospheric offsets and the consequent sensitivity degradation, suitable consideration of maximum baselines and the expected ionospheric offsets, can help to make more optimal (effective) use of the full uGMRT for sensitive pulsar observations within its Band 2 range.

Beyond the currently operational low-frequency facilities, the upcoming SKA-Low will also necessarily benefit from such considerations. A significant subset of pulsar science planned with the SKA (in particular those involving timing or single-pulse studies) rely on sub-arraying, and hence involve sub-grouping of stations extending out to baselines of $\sim$10 km. While the much higher sensitivity offered by SKA-Low will readily allow optimal sub-grouping of stations, considerations along the lines discussed here will likely become important for maximising achievable sensitivity, especially for beamformed observations at frequencies $\lesssim$ 150 MHz. For instance, high-sensitivity observations in this lower SKA-Low band are likely to benefit from sub-grouping of stations within an extent of $\lesssim$ 1-2 km, which may not be possible for stations located in the outer parts of the array. For these outer core stations, suitable sub-grouping within $\lesssim$2-3 km may help mitigate the ionospheric effects, while any sub-grouping involving stations with $\gtrsim$5 km baselines may require mitigation schemes similar to those discussed here, especially given that SKA-Low is to be built at the same site as the MWA, and so ionospheric effects will be quite similar.

\section{SUMMARY}

The multi-pixel beamformer is a factor $\sim$10 more efficient than previous MWA beamformer iterations without affecting the sensitivity of pulsar detections. The portability of the software has been proven by installing it on three supercomputers, which can share the processing load of large-scale surveys between multiple institutions. These improvements make it feasible to perform large scale pulsar surveys with the MWA.

We investigated the ionosphere's effect on MWA VCS observations and characterised them as the bulk and residual offsets. The ionospheric residual offsets only affect sensitivity below 140 MHz when the ionosphere is very turbulent. The bulk ionospheric offsets reduce sensitivity when the bulk offset differs between the calibration and target observation, as illustrated in Figure \ref{fig:bulkoffset}. This bulk offset error can be measured and corrected using the method described in \S \ref{sec:offsetfix}. Correcting this bulk offset makes our pulsar candidate position estimates more accurate, and our improved understanding of the ionosphere provides more realistic position uncertainties. Mitigating the ionospheric offsets will become more important for MWA Phase III and should be considered in the design of SKA-Low.

\section*{Acknowledgement}
%MRO
This scientific work makes use of the Murchison Radio-astronomy Observatory, operated by CSIRO. We acknowledge the Wajarri Yamatji people as the traditional owners of the Observatory site.
%Pawsey
This work was supported by resources provided by the Pawsey Supercomputing Centre with funding from the Australian Government and the Government of Western Australia.
%OzSTAR
This work was supported by resources awarded under Astronomy Australia Ltd's ASTAC merit allocation scheme on the OzSTAR national facility at the Swinburne University of Technology. The OzSTAR program receives funding in part from the Astronomy National Collaborative Research Infrastructure Strategy (NCRIS) allocation provided by the Australian Government.
% Reviewer
We thank the referee for several useful comments that improved the presentation and clarity of this paper.

\bibliography{mendeley_references}

\end{document}